\documentclass[]{aa}

\usepackage{txfonts}
\usepackage{graphicx}
\usepackage{hyperref}
\usepackage{xcolor}
\usepackage{algorithm}
\usepackage{algorithmicx}
\usepackage[inline]{enumitem}
\usepackage{natbib}
\bibpunct{(}{)}{;}{a}{}{,}

\hypersetup{colorlinks=true,citecolor=blue,linkcolor=blue}

\begin{document}

\title{
    Formation of flattened planetesimals by gravitational collapse of rotating pebble clouds
    }

\author{
    Sebastian~Lorek\inst{1},
    Anders~Johansen\inst{1,2}
    }

\institute{
    Centre for Star and Planet Formation,
    Globe Institute, 
    University of Copenhagen,
    {\O}ster Voldgade 5–7, 
    DK-1350 Copenhagen, 
    Denmark \\
    \email{sebastian.lorek@sund.ku.dk}
    \and
    Lund Observatory, 
    Department of Astronomy and Theoretical Physics, 
    Lund University,
    Box 43, 
    221 00 Lund,
    Sweden
    }

\date{Received ; accepted }

\abstract{
Planetesimals are believed to form by the gravitational collapse of aerodynamically concentrated clumps of pebbles. Many properties of the objects in the cold classical Kuiper belt -- such as binarity, rotation, and size distribution -- are in agreement with this gravitational collapse model. Further support comes from the pebble-pile structure inferred for comet nuclei. For this study, we simulated the final assembly of a planetesimal from the gravitational collapse of a rotating clump of pebbles. We implemented a numerical method from granular dynamics to follow the collapse that includes the transition from a pebble swarm to solid cells at a high density. We compared the shapes of the simulated planetesimals with the shapes of the lobes of contact binaries and bilobed Solar System objects. We find that the gravitational collapse of slowly rotating pebble clouds naturally explains the formation of flattened ellipsoidal bodies. This result agrees well with the flattened structure of the bilobed planetesimal Arrokoth and the shapes of the components of bilobed comets.
}

\keywords{
    Methods: numerical --
    Planets and satellites: formation
    }

\maketitle


\section{Introduction}
\label{sec:introduction}

Planetesimals are believed to form by the gentle gravitational collapse of aerodynamically concentrated clumps of pebbles (pebble clouds). Pebbles are millimetre- to decimetre-sized aggregates that have grown by collisions of micrometre-sized dust and ices grains in the gaseous environment of protoplanetary discs. Laboratory and numerical studies of collisional dust growth have shown that bouncing, erosion, fragmentation, and radial drift limit the maximum sizes to which pebbles can grow to the millimetre- to centimetre-size range \citep{Blum2008,Guettler2010,Zsom2010,Krijt2015,Blum2018,Lorek2018}. The streaming instability operates well for these pebble sizes, a process that has been demonstrated numerically to be efficient in concentrating large amounts of pebbles into filament-like structures that subsequently fragment under their own gravity to form pebble-pile planetesimals \citep{Youdin2005,Johansen2007,Johansen2014,WahlbergJansson2014}.

The minor bodies in the Solar System, such as asteroids, comets, and Kuiper belt objects, are remnants of the planetesimals that formed through this process during the formation of the Solar System about $4.5$ billion years ago. Therefore, the minor body populations provide a glimpse into the past and offer a unique possibility to study the properties of planetesimals. Minor bodies, especially the ones that are not massive enough for self-gravity to press them into spheres, often have irregular shapes which might be reminiscent of their formation or caused by evolutionary processes throughout the history of the Solar System.

The main asteroid belt between the orbits of Mars and Jupiter is a collisionally evolved minor body population and only objects with diameters larger than ${\sim}120\,\mathrm{km}$ are thought to be primordial, which means that the properties of the large objects were shaped during the accretion process, while the smaller asteroids are by-products of collisional fragmentation \citep{Bottke2005}. The largest asteroids, for example Ceres and Vesta, are massive enough to be spherical. Smaller asteroids are rubble piles that formed through gravitational re-accumulation of the fragments from catastrophic collisions of larger bodies \citep{Michel2013,Michel2020}. Therefore, the irregular shapes of kilometre-sized asteroids are not primordial and not related to the planetesimal formation process.

Comets formed outside the water ice line where, besides refractory dust, volatiles were also present as ices. The favoured formation scenario for comets is the gravitational collapse of a pebble cloud, which is consistent with the observed properties of cometary nuclei \citep{Blum2014,Blum2017,Blum2022}. From spacecraft flybys it is known that cometary nuclei are irregularly shaped, many are elongated or bilobed, with sizes in the range of kilometres to a few tens of kilometres \citep{Keller1986,Buratti2004,Duxbury2004,AHearn2005,AHearn2011,Sierks2015}. When entering the inner Solar System, solar irradiation triggers the sublimation of volatile ices and activity-driven erosion changes the surface morphology of the nuclei. Anisotropic mass loss caused by non-uniform insolation of the nucleus is capable of producing irregular-shaped nuclei \citep{Vavilov2019}. While transitioning from the comet reservoir to the inner Solar System, sublimation of volatiles can spin up the nucleus to disruption and the subsequent reassembly of the fragments can also lead to a bilobed nucleus \citep{Safrit2021}. Furthermore, modelling the collisional history of comet-sized planetesimals during the dynamical evolution of the outer Solar System suggests that kilometre-sized cometary nuclei are not primordial but predominantly the fragments of larger bodies \citep{Morbidelli2015}, although the collision history depends very strongly on the unknown lifetime of the primordial, massive Kuiper belt before the outward migration of Neptune. Numerical simulations have demonstrated that the re-accumulation of  the fragments produced in sub-catastrophic or catastrophic collisions would lead to the elongated or bilobed shapes of cometary nuclei while preserving volatiles and porosity \citep{Jutzi2017,Schwartz2018}. This shows that the shapes of cometary nuclei as observed today are not necessarily primordial, but can originate from evolutionary processes as well.

Kuiper belt objects (KBOs) orbiting in the region beyond Neptune are another population of minor bodies in the Solar System. The inclination distribution of the Kuiper belt objects shows that there are two populations, a hot population with a wide inclination distribution and a cold population with a narrow inclination distribution \citep{Brown2001}. The transition between both populations is set around $5^\circ$ and KBOs that have an inclination ${\lesssim}5^\circ$ belong more likely to the cold population. This cold population, the so-called cold classical Kuiper belt, forms a narrow ring between approximately $42.4\,\mathrm{au}$ and $47.7\,\mathrm{au}$, with a sub-population centred at ${\sim}44\,\mathrm{au}$ (the so-called kernel), and consists most likely of planetesimals that formed in situ, unaffected by the dynamical history of the outer Solar System formation \citep{Batygin2011,Nesvorny2011,Petit2011,Kavelaars2021}.

\citet{Kavelaars2021} compared the size distributions of the cold classical Kuiper belt with the size distributions typically found in streaming instability simulations. They find a good match with the exponentially tapered power law proposed in \citet{Schaefer2017}. However, the total mass of the classical Kuiper belt is at least a factor 10 lower than the result of streaming instability; this may be due to gravitational perturbations from Neptune \citep{Gomes2018}. We refer readers to \citet{Simon2022} for a detailed discussion of the streaming instability initial-mass function compared to the classical Kuiper belt.

Observations show that the binary fraction among the cold classical Kuiper belt objects is very high \citep{Noll2008b,Noll2008} and it is even hypothesised that almost all objects of the Kuiper belt formed as binaries \citep{Fraser2017}. The high binary fraction and the mostly prograde mutual orbits of the binaries are in agreement with high-resolution studies of the gravitational collapse of pebble clouds \citep{Nesvorny2010,Nesvorny2019,Robinson2020,Nesvorny2021,Polak2023}. Light curve modelling of KBOs revealed that besides wide binaries, contact binaries are also present among the KBOs. From the first confirmed contact binary 2001~QG$_{298}$, a contact-binary fraction of ${\gtrsim}10{-}30\,\%$ was estimated \citep{Sheppard2004,Lacerda2011}. Other candidates are the Kuiper belt objects 2003~SQ$_{317}$ and 2004~TT$_{357}$ that have light curves consistent with those of contact binaries even though highly elongated shapes cannot be entirely ruled out \citep{Lacerda2014,Thirouin2017}. The contact-binary fraction among the Plutinos, which are KBOs such as Pluto that are in a 3:2 resonance with Neptune, is even higher with an estimate of ${\sim}50\,\%$ based on observations \citep{Thirouin2018}. This shows that contact binaries are frequent with a fraction of $10{-}25\,\%$ among the cold classical Kuiper belt objects and ${\sim}50\,\%$ among the Plutinos \citep{Thirouin2019,Showalter2021}. The cold classical Kuiper belt object (486958)~Arrokoth is a contact binary that was visited by the National Aeronautics and Space Administration (NASA) spacecraft New Horizons \citep{Stern2019}. As a member of the cold classical Kuiper belt and a contact binary, Arrokoth most likely represents a primordial planetesimal that has formed by the gravitational collapse of a pebble cloud \citep{Stern2019,McKinnon2020}. Even though it is hypothesised that the shape of Arrokoth might be the result of volatile sublimation on a $1{-}100\,\mathrm{Myr}$ timescale \citep{Zhao2021}, its shape might still provide useful insights in the primordial shapes of planetesimals.

While the small asteroids are the results of catastrophic collisions and gravitational re-accumulation, comets and the cold classical Kuiper belt objects might still be representative of the shapes of the kilometre-sized primordial planetesimals that formed through the gravitational collapse of pebble clouds. The dynamics of the pebble cloud collapse has been studied in high resolution $N$-body simulations or by means of hydrodynamic simulations of the self-gravitating pebble fluid \citep{Nesvorny2010,Nesvorny2019,Robinson2020,Nesvorny2021,Polak2023}. However, these studies do not resolve the final shapes of the planetesimals. Here, we attempt to model the final assembly of a kilometre-sized planetesimal from the gravitational collapse of a rotating clump of pebbles using a Monte-Carlo method for granular dynamics and show that, depending on the angular momentum content, the final shapes of kilometre-sized planetesimals are in agreement with the shapes of the left-over planetesimals in the Solar System.

The paper is organised as follows. In Section~\ref{sec:shapesofsolarsystembodies}, we broadly summarise available shape information of minor bodies. In Section~\ref{sec:methods}, we provide an overview of the Monte-Carlo method, with a more detailed description in the Appendix~\ref{sec:montecarlomethod}. In Section~\ref{sec:simulatingthefinalassembly}, we go throught the results of our simulations. In Section~\ref{sec:discussion}, we discuss our results in the context of the Solar System bodies. Finally, in Section~\ref{sec:conclusions}, we summarise the main findings of our study.


\section{Shapes of minor bodies}
\label{sec:shapesofsolarsystembodies}

In this section, we present a brief overview of minor bodies with well-studied sizes, namely the small number of comets that were visited by space craft, the cold classical Kuiper belt object Arrokoth, and the interstellar asteroid 1I/2017 U1 (`Oumuamua).

\begin{table*}
\caption{Comet and KBO sizes.}
\label{tab:cometandKBOsizes}
\centering
\begin{tabular}{lcccccccl}
\hline\hline
Object & $a$ & $b$& $c$ & $b/a$ & $c/a$ & $c/b$ & Fit & Ref \\
& (km) & (km) & (km) & & & & \\
\hline
1P/Halley & $15.30$ & $7.22$ & $7.20$ & $0.472$ & $0.471$ & $0.997$ & shape & 1 \\
8P/Tuttle &  &  & &  &  &  & &  \\
\quad big lobe & $5.75$ & $4.11$ & $4.11$ & $0.715$ & $0.715$ & $1.000$ & prolate  & 2 \\
\quad small lobe & $4.25$ & $3.27$ & $3.27$ & $0.769$ & $0.769$ & $1.000$ & prolate  & 2 \\
9P/Tempel 1 & $7.6$ & $4.9$ & $4.9$ & $0.645$ & $0.645$ & $1.000$ & shape & 3 \\
19P/Borrelly & $8.00$ & $3.16$ & $3.16$ & $0.395$ & $0.395$ & $1.000$ & prolate & 4 \\
67P/Churyumov-Gerasimenko & $4.34$ & $2.60$ & $2.12$ & $0.599$ & $0.488$ & $0.815$ & shape & 5 \\
\quad big lobe & $4.10$ & $3.52$ & $1.63$ & $0.859$ & $0.398$ & $0.463$ & shape & 5 \\
\quad small lobe & $2.50$ & $2.14$ & $1.64$ & $0.856$ & $0.656$ & $0.766$ & shape & 5 \\
81P/Wild 2 & $5.5$ & $4.0$ & $3.3$ & $0.727$ & $0.600$ & $0.825$ & ellipsoid & 6 \\
103P/Hartley 2 & $2.33$ & $0.69$ & $0.69$ & $0.296$ & $0.296$ & $1.000$ & shape & 7 \\
1I/2017 U1 (`Oumuamua) & & & & & & & & \\
\quad DISC & $0.115$ & $0.111$ & $0.019$ & $0.965$ & $0.165$ & $0.171$ & ellipsoid & 9 \\
\quad CIGAR & $0.324$ & $0.042$ & $0.042$ & $0.130$ & $0.130$ & $1.000$ & ellipsoid & 9  \\
(486958) Arrokoth & $35.95$ & $19.90$ & $9.75$ & $0.539$ & $0.264$ & $0.490$ & shape & 8 \\
\quad big lobe & $21.20$ & $19.90$ & $9.05$ & $0.939$ & $0.427$ & $0.455$ & shape & 8 \\
\quad small lobe & $15.75$ & $13.85$ & $9.75$ & $0.879$ & $0.619$ & $0.704$ & shape & 8 \\
\hline
\end{tabular}
\tablefoot{Axis dimensions $a$, $b$, and $c$ are diameters. Fit refers to how the axis dimensions were estimated: shape model (shape), triaxial ellipsoid fit (ellipsoid), or prolate ellipsoid fit (prolate). For prolate models, the two small axes $b$ and $c$ are equal. A shape model is a 3D reconstruction of the shape of the object based on images taken by the spacecraft.}
\tablebib{
(1) \citet{Merenyl1990}; (2) \citet{Harmon2010}; (3) \citet{AHearn2005}; (4) \citet{Buratti2004}; (5) \citet{Jorda2016}; (6) \citet{Duxbury2004}; (7) \citet{AHearn2011}; (8) \citet{Keane2022}; (9) \citet{Mashchenko2019}.}
\end{table*}

\subsection{Arrokoth}
The overall dimensions of Arrokoth are $35.95{\times}19.90{\times}9.75\,\mathrm{km}$. Because the flyby of the New Horizons spacecraft was fast and distant and because Arrokoth lacks any natural satellites, there are no direct constraints on the mass or the density by gravity measurements \citep{Keane2022}. The density is hence inferred from the geophysical environment of Arrokoth which favours a nominal density of ${\sim}235\,\mathrm{kg}\,\mathrm{m}^{-3}$ and yields a mass of ${\sim}7.485{\times}10^{14}\,\mathrm{kg}$ for the entire body \citep{Keane2022}. The two planetesimals that form the lobes of Arrokoth have dimensions of $21.20{\times}19.90{\times}9.05\,\mathrm{km}$ for the large lobe Wenu and $15.75{\times}13.85{\times}9.75\,\mathrm{km}$ for the small lobe Weeyo \citep{Keane2022}. The large lobe is hence significantly flattened along one axis, while the small lobe is slightly less flat. It is hypothesised that the flattening could be the result of volatile sublimation on a $1{-}100\,\mathrm{Myr}$ timescale \citep{Zhao2021}. With a bulk density of ${\sim}235\,\mathrm{kg}\,\mathrm{m}^{-3}$, Arrokoth is less dense than other minor bodies of the Solar System and especially less dense than comets that have an estimated mean density of around $500\,\mathrm{kg}\,\mathrm{m}^{-3}$. For a typical grain density of $2154\,\mathrm{kg}\,\mathrm{m}^{-3}$ (rock mass-fraction of $75\,\%$ with an average rock density of $3500\,\mathrm{kg}\,\mathrm{m}^{-3}$), the volume-filling fraction of Arrokoth is only $11\,\%$, which translates to a porosity of $89\,\%$ \citep{Keane2022}. As a member of the cold classical Kuiper belt and a contact binary, Arrokoth most likely represents a primordial planetesimal that has formed through the gravitational collapse of a pebble cloud and the subsequent low-velocity merger of two components \citep{Stern2019,McKinnon2020,Marohnic2021}. Therefore, the shape of Arrokoth provides useful insights in the primordial shapes of planetesimals.

\subsection{Comets}
Because of their small sizes, cometary nuclei are difficult to observe from Earth. However, information about their sizes and shapes can be obtained through light curve analysis or radar observation. Additionally, for six comets that were visited by spacecraft, detailed information about the shapes of their nuclei exists. The Giotto mission of the European Space Agency (ESA) to comet 1P/Halley obtained for the first time a direct view of a cometary nucleus and revealed an irregular-shaped kilometre-sized object \citep{Keller1986,Keller1987}. Further NASA missions to comets 19P/Borrelly (Deep Space 1), 9P/Tempel~1 and 103P/Hartley~2 (Deep Impact), and 81P/Wild~2 (Stardust) provided more examples for the irregular shapes of cometary nuclei \citep{Buratti2004,Duxbury2004,AHearn2005,AHearn2011}. Lastly, the most recent mission to a comet, ESA's Rosetta mission, which accompanied the Jupiter-family comet 67P/Churyumov-Gerasimenko for almost two years along its orbit, obtained high-resolution images and detailed shape information of a kilometre-sized bilobed nucleus that is thought to be a contact binary of two distinct lobes \citep{Massironi2015,Nesvorny2018}, with a large lobe of size $4.1{\times}3.5{\times}1.6\,\mathrm{km}$ and a small lobe of size $2.5{\times}2.1{\times}1.6\,\mathrm{km}$ \citep{Sierks2015,Jorda2016}. The two lobes are not spherical but flattened along one axis. The nucleus of 67P has a mass of $9.982{\times}10^{12}\,\mathrm{kg}$, a density of $532\,\mathrm{kg}\,\mathrm{m}^{-3}$, and a high porosity of the order of $70\,\%$ \citep{Jorda2016}. Shape models are also available for comets 103P/Hartley~2 and 19P/Borrelly. Both comets are bilobed and it is speculated that this is the result of formation from individual objects \citep{Oberst2004,Thomas2013}. Radar observations of comet 8P/Tuttle revealed a contact binary with a large lobe of size $5.75{\times}4.11\,\mathrm{km}$ and a small lobe of size $4.25{\times}3.27\,\mathrm{km}$ \citep{Harmon2010}.

\subsection{1I/2017 U1 (`Oumuamua)}
The object 1I/2017 U1 (`Oumuamua) was the first detected interstellar object to pass through the Solar System \citep{Meech2017}. Light curve analysis and shape modelling revealed that `Oumuamua is unusually elongated with an axes ratio as high as 6:1 suggesting an ellipsoid body with axes diameters of $230{\times}35{\times}35\,\mathrm{m}$ \citep{Jewitt2017}. The overall best-fit model of \citet{Mashchenko2019} suggests a thin disc with dimensions $115{\times}111{\times}19\,\mathrm{m}$. A cigar-shaped body with dimensions of $324{\times}42{\times}42\,\mathrm{m}$ is also possible, but less likely according to the analysis of \citet{Mashchenko2019}, as the thin disc model fits the observed light-curve minima best.


\section{Methods}
\label{sec:methods}

To simulate the shape of a planetesimals, it is necessary to study the gravitational collapse of a pebble cloud. However, even a kilometre-sized planetesimal requires ${\sim}10^{18}$ millimetre-sized pebbles to form. It is hence computationally impossible to do this directly by following each pebble individually. However, one can simulate the time evolution of the distribution function of the pebbles subject to collisions and self gravity, or in other words, solve the Boltzmann equation. We do this here by using a direct-simulation Monte Carlo (DSMC) method \citep{Bird1994,Alexander1997,Poeschel2005}. The key idea of DSMC is to sample the distribution function with a number of computational particles, each of which represents a swarm of physical pebbles with equal properties. The Boltzmann equation, which governs the time evolution of the distribution function, is then split into an advection step, in which the positions of the particles are propagated with their current velocities, and a collision step, in which the particle velocities change due to collisions (see Appendix~\ref{sec:montecarlomethod} for a detailed description). Self gravity is accounted for in the advection step by using a drift-kick-drift scheme and a Poisson solver based on the Fast Fourier Transformation (FFT) (see Appendix~\ref{sec:selfgravityofthecloud}).

The collision treatment of the DSMC method based on the Boltzmann equation was originally developed to simulate dilute gases. In this case, the assumption is that two particles move independently from each other and that the two-particle distribution function can be written as the product of two one-particle distribution functions
\begin{equation}
f_2(\mathbf{x}_1,\mathbf{x}_2,\mathbf{v}_1,\mathbf{v}_2,t)=f(\mathbf{x}_1,\mathbf{v}_1,t)f(\mathbf{x}_2,\mathbf{v}_2,t)
\end{equation}
 (molecular chaos hypothesis). For dense gases, however, this assumption may not hold any more because of correlations between the particles due to their finite volumes. The most simple correction to account for the finite volume effects is to introduce the Enskog factor $\chi$, which is basically the two-particle correlation function, such that the distribution function reads
\begin{equation}
f_2(\mathbf{x}_1,\mathbf{x}_2,\mathbf{v}_1,\mathbf{v}_2,t)=\chi f(\mathbf{x}_1,\mathbf{v}_1,t)f(\mathbf{x}_2,\mathbf{v}_2,t)
\end{equation}
\citep{Poeschel2005}. The Enskog factor $\chi$ depends on the local density of particles and can be derived from the equation of state of a hard-sphere fluid \citep{Ma1986} (see Appendix~\ref{sec:montecarlomethod}). The Boltzmann equation then turns into the Boltzmann-Enskog equation, which is the basis for our study \citep{Montanero1996,Montanero1997}.

To spatially resolve the planetesimal, space is discretised using a Cartesian grid. While the Enskog correction makes sure that the collision rates of the DSMC method are correct, it does not affect the propagation of particles. In spatially resolved simulations, particles hence move unconditionally and unaffected by other particles, which can lead to an unrealistically high packing of particles within cells. To prevent this, the propagation step is modified and new positions are accepted based on the local packing of particles \citep{Hong2021}. This probabilistic approach prevents the over-packing of cells and further allows to identify fully packed cells as the solid elements comprising the planetesimal. The maximum volume-filling factor of a cell is ${\sim}0.64$, which is achieved for a random close packing of hard spheres.

A significant limitation of the method is that because of the fixed grid, it is not possible to properly resolve rotation and hence binary formation as once cells fill up, the bulk of the particles remains in the cell as the mean-free path is so short that only particles close to the cell boundaries might be able to diffuse outwards. However, for clumps of low angular momentum that collapse into a single body, the method is able to reproduce the shapes of planetesimals.


\section{Simulating the final assembly}
\label{sec:simulatingthefinalassembly}

We focus on the final assembly of a kilometre-sized planetesimal from a pebble cloud with low angular momentum content. We cannot resolve the entire collapse from a Hill-radius-sized pebble cloud down to a planetesimal because the required number of grid cells to achieve sub-kilometre resolution would be too high. Instead, we assume that the pebble cloud has already fragmented into sub-clumps that would eventually form, for example, the components of a binary as has been shown in $N$-body simulations of the gravitational collapse \citep[e.g.][]{Polak2023}.

\subsection{Initial conditions}

We started all simulations with an initial radius of the pebble cloud that was $10$ times larger than the spherical planetesimal size corresponding to the total mass. The simulation box and the number of grid cells were chosen such that we achieved a resolution of $1/10$ of the planetesimal radius. For a planetesimal of radius $R_\mathrm{p}{=}1\,\mathrm{km}$ (as chosen in this study) this meant that the initial cloud radius was $R_\mathrm{c}{=}10\,\mathrm{km}$ and the resolution was $0.1\,\mathrm{km}$. The pebbles had a fixed radius of $0.5\,\mathrm{mm}$ in agreement with expected pebble sizes \citep{Zsom2010,Lorek2018} and a constant coefficient of restitution of $0.5$ \citep{Weidling2015}. The pebbles were initially uniformly distributed within the cloud volume and their initial velocities were isotropic following a Maxwellian distribution with $\sqrt{k_\mathrm{B}T/m}{\approx}(1/3)v_\mathrm{vir}$, where $v_\mathrm{vir}$ is the virial velocity of the pebble cloud. In addition, we added rotation to the pebble cloud by setting an initial angular momentum $L$. A critically rotating Jacobi ellipsoid of mass $M_\mathrm{p}$ and effective radius $R_\mathrm{p}$ has an angular momentum of
\begin{equation}
L_\mathrm{J}{=}0.4\sqrt{GM_\mathrm{p}^3R_\mathrm{p}}.
\end{equation}
We therefore set the angular momentum content in units of $L_\mathrm{J}$ because the cloud can collapse into a single body only for $L/L_\mathrm{J}{\lesssim}1$. To study the effect of angular momentum on the planetesimal shape we systematically varied $L/L_\mathrm{J}$ between $0$ (no rotation) and $2$ (super-critical rotation).

We conducted the study for a pebble cloud that had the mass of a planetesimal of $R_\mathrm{p}{=}1\,\mathrm{km}$ and density $1280\,\mathrm{kg}\,\mathrm{m}^{-3}$, which corresponded to a volume-filling factor of the planetesimal of $0.64$ and a material density of the pebbles of $2000\,\mathrm{kg}\,\mathrm{m}^3$. The pebble cloud was considered in isolation and therefore its dynamics was independent of the heliocentric distance and rotation around the Sun. This is a reasonable assumption because the free-fall time for such a small cloud is of the order of days, which is much shorter than the orbital period in the Kuiper belt, and furthermore, the size of the cloud is significantly smaller than the Hill radius so that shearing effects can be neglected.

To identify a planetesimal, we ran the simulations for typically ${\sim}1.2{-}1.4$ free-fall times of the cloud until a solid body formed that no longer changed its shape significantly. The free-fall time was of the order of a day. We then identified all cells that had a volume-filling factor ${\ge}0.5$, which is the loosest stable packing of spheres \citep{Torquato2007}. Because we considered the pebbles as spheres and because we did not have a size distribution of pebbles, we set the maximum volume-filling factor that a cell could obtain to $0.64$, which is obtained for a random close packing of spheres and slightly higher than that for a random loose packing of spheres in the zero-gravity limit with volume-filling factor of $0.56$ \citep{Onoda1990}.

\subsection{Planetesimal shapes}

\begin{figure*}
\centering
\includegraphics[width=17cm]{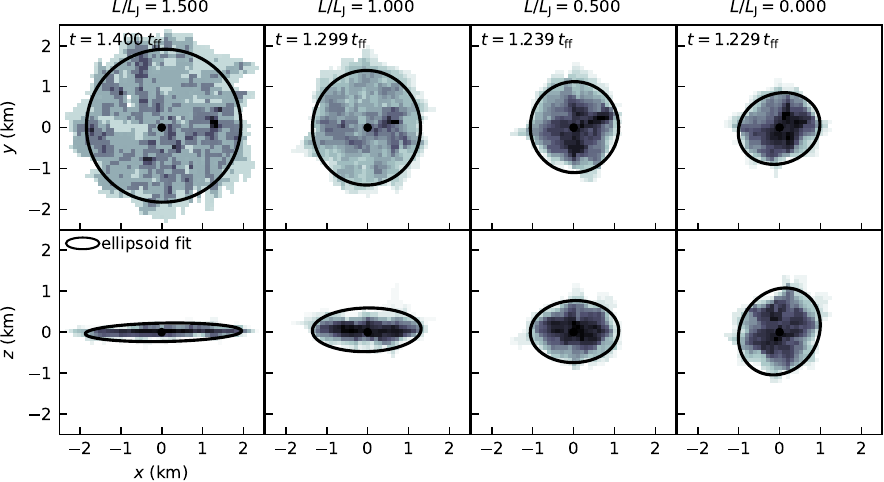}
\caption{Shapes of planetesimals for different values of the initial angular momentum. The rows show the column densities of cells that have a volume-filling factor ${\ge}0.5$ in the $xy$- and $xz$-planes for different values of the initial angular momentum of the cloud at the end of each simulation. We omit the $yz$-plane because of the evident cylindrical symmetry. The contour of the ellipsoid fitted to the planetesimals in the respective plane are shown in black with the centre of the ellipse marked with a black circle. The figure shows the results for initial values of $L/L_\mathrm{J}{=} 1.5$, $1.0$, $0.5$ and $0.0$ to illustrate the effect of angular momentum of the cloud on the flattening of the planetesimals.}
\label{fig:density_overview_compact}
\end{figure*}

Figure~\ref{fig:density_overview_compact} shows a selection of planetesimals that formed from clouds with different initial angular momentum. With increasing angular momentum of the cloud, the planetesimals flatten in the $z$-direction which is parallel to the initial rotation axis. In the plane perpendicular to the rotation axis ($xy$-plane), the planetesimals remain more round. As expected for very high angular momentum content $L/L_\mathrm{J}{\gtrsim}1.5$, the cloud does not collapse into a single object but instead forms a flat disc with spiral structures. Those clouds would possibly form binary systems. For lower angular momentum in the range $1{\lesssim}L/L_\mathrm{J}{\lesssim}1.5$, very thin disc-like objects form. When the angular momentum is $L/L_\mathrm{J}{\lesssim}1$, the planetesimal obtain lenticular and nearly spherical shapes comparable to the shapes of `Oumuamua or the lobes of Arrokoth and 67P (see discussion in Sect.~\ref{sec:discussion}). In summary, we find disc-like and lenticular-shaped planetesimals in our simulations. However, we do not see elongated ellipsoids that would resemble bodies such as 19P/Borrelly or 103P/Hartley~2.

\subsection{Axes ratios of planetesimals}

We used linear least-squares to fit an ellipsoid that is centred at the centre-of-mass of the planetesimal to the boundary points (see Appendix~\ref{sec:ellipsoidalfit}). The orientation of the ellipsoid does not necessarily coincide with the Cartesian axes and we hence determined the eigenvalues and eigenvectors of the ellipsoid which give the lengths and the directions of the main axes. Because of the limited resolution of the Cartesian grid, the ellipsoid is only an approximation to quantify the (irregular) shape of the planetesimal. To check the validity of the approximation, however, we calculated the inertia tensor of the planetesimal and compared the directions of the principal axes with the main axes of the ellipsoid. The axes were aligned, which showed that the ellipsoid orientation matched the mass distribution of the body. We quantify the shape of the planetesimals by calculating the ratios of the three axes of the fitted ellipsoid. We labelled the axes $a$, $b$, and $c$ such that $c{\leq}b{\leq}a$ and calculated the ratios $b/a$, $c/a$, and $c/b$. The closer any of the ratios is to unity, the more spherical is the object. For example, a planetesimal with $b/a{=}1$, but $c/a{=}0.5$ and $c/b{=}0.5$, would be round in the $ba$-plane and flattened in the $ca$- and $cb$-planes resulting in a lenticular shape.

\begin{figure}
\resizebox{\hsize}{!}{\includegraphics{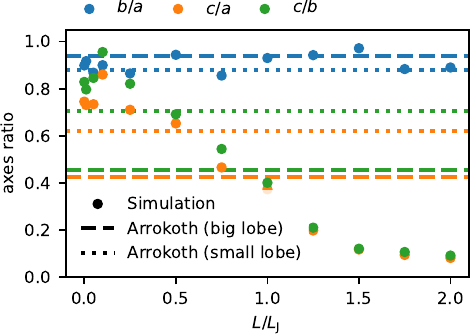}}
\caption{Axes ratios of planetesimals as a function of the initial angular momentum of the cloud. The axes ratios of the two lobes of Arrokoth are shown as dashed (big lobe) and dotted (small lobe) lines. }
\label{fig:aspect_ratio}
\end{figure}

\begin{figure}
\resizebox{\hsize}{!}{\includegraphics{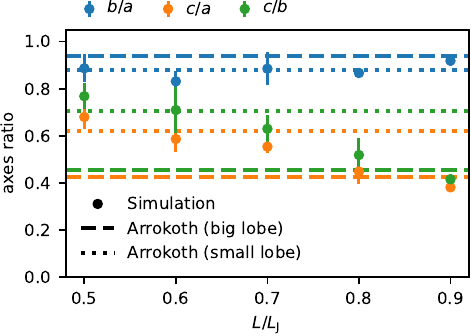}}
\caption{Average axes ratios of planetesimals for the initial angular momentum of the cloud in the range $0.5$ to $0.9$. For each initial $L/L_\mathrm{J}$, the results of $5$ different realisations are averaged. The axes ratios of the two lobes of Arrokoth are shown as dashed (big lobe) and dotted (small lobe) lines.}
\label{fig:aspect_ratio_mean}
\end{figure}

Figure~\ref{fig:aspect_ratio} shows the axes ratios of the planetesimals as a function of the initial angular momentum content. The figure confirms that for $L/L_\mathrm{J}{\lesssim}1$ the planetesimals become more spherical as is already indicated in Figure~\ref{fig:density_overview_compact}. For comparison, we plot the axes ratios for the two lobes of Arrokoth (see Table~\ref{tab:cometandKBOsizes} for the shape information). Our model is consistent with the axes ratios, and hence the shapes of the two planetesimals forming Arrokoth, for pebble clouds with angular momentum approximately in the range $0.5{\lesssim}L/L_\mathrm{J}{\lesssim}0.9$. Here, we chose Arrokoth for comparison because being a cold classical Kuiper belt object, Arrokoth most likely represents a primordial planetesimal. Figure~\ref{fig:aspect_ratio_mean} shows the average axes ratios for planetesimals that formed from clumps with initial angular momentum $L/L_\mathrm{J}$ in the range $0.5{-}0.9$. For each value of $L/L_\mathrm{J}$, we averaged the axes ratios over five realisations for the same initial conditions to obtain a measure for the uncertainty of our Monte-Carlo simulations. Figure~\ref{fig:aspect_ratio_mean} shows that the uncertainty of the axes ratios are ${\sim}10\,\%$.

\subsection{Angular momentum}

\begin{figure}
\resizebox{\hsize}{!}{\includegraphics{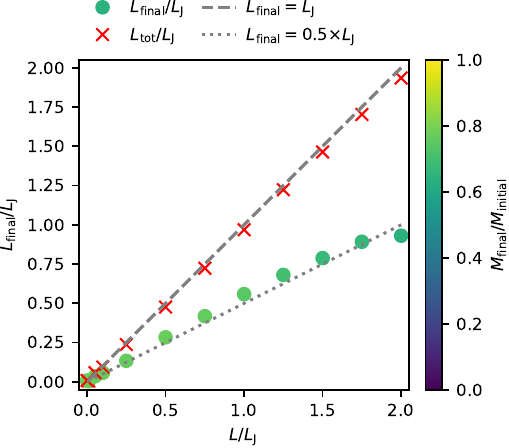}}
\caption{Angular momentum of planetesimals as a function of the initial angular momentum of the cloud. The dashed line indicates where the final angular momentum of the planetesimal $L_\mathrm{final}$ equals the initial angular momentum and the dotted line where $L_\mathrm{final}$ equals half of the initial value. The colour coding corresponds to the mass that ends up in the planetesimal. The red crosses show the total angular momentum $L_\mathrm{tot}$ in the simulation.}
\label{fig:angmom}
\end{figure}

Figure~\ref{fig:angmom} shows the final angular momentum of the planetesimals as a function of the initial angular momentum of the pebble cloud. We find that the final planetesimal contains about $50\,\%$ of the initial angular momentum and that between $60{-}80\,\%$ of the mass of the pebble cloud ends up in the planetesimal. The figure also shows that the total angular momentum is conserved within ${\sim}5\,\%$. This deviation is the result of the open boundary conditions and the loss of particles that leave the simulation box taking away a tiny fraction of the total angular momentum.


\section{Discussion}
\label{sec:discussion}

In this section we discuss our results in the context of the minor bodies of the Solar System. We provided a brief overview of the processes that shape various minor bodies in the Solar System in Section~\ref{sec:introduction}. Small asteroids are typically rubble piles that formed through the gravitational reassembly of fragments produced in catastrophic collisions of larger bodies \citep{Michel2013,Michel2020}. The shapes of kilometre-sized asteroids are hence most likely not primordial and not related to the gravitational collapse of pebble clouds. We therefore exclude asteroids from our discussion. Cometary nuclei have irregular shapes that can originate in their formation process or be the result of evolutionary process \citep[e.g.][]{Jutzi2015,Jutzi2017,Schwartz2018,Safrit2021}. Especially the shapes of the comets that are clearly contact binaries might provide some information about the formation in collapsing pebble clouds. Lastly, we include Arrokoth in our discussion. As a cold classical Kuiper belt objects, Arrokoth resembles best a primordial planetesimal \citep{Stern2019,McKinnon2020}. Table~\ref{tab:cometandKBOsizes} summarises the available shape information for Solar System bodies that we use in our comparison.

\subsection{Comparison with simulations}

\begin{figure*}
\centering
\includegraphics[width=17cm]{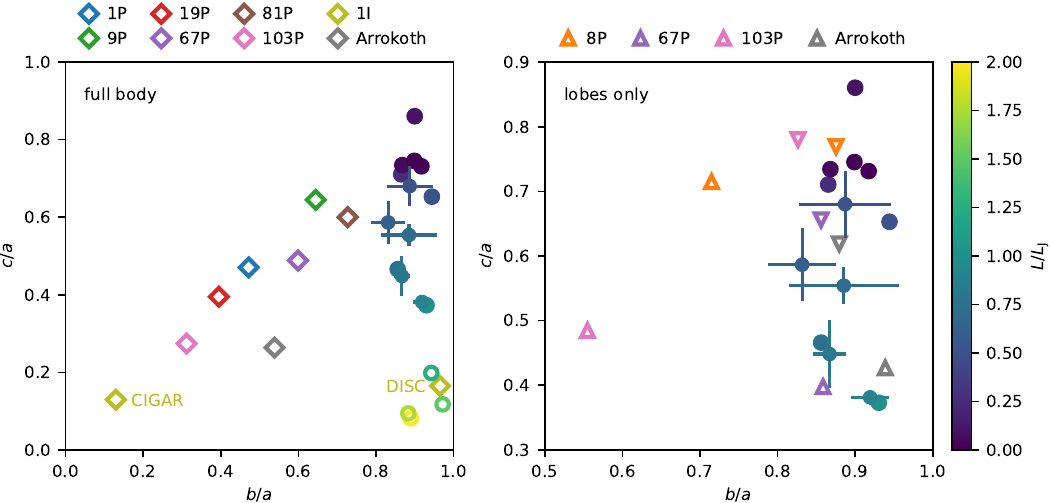}
\caption{Axes ratios of Solar System bodies, `Oumuamua, and simulated planetesimals. The simulated planetesimals are shown as circles with a colour that indicates the initial angular momentum of the pebble cloud $L{/}L_\mathrm{J}$. For $L{/}L_\mathrm{J}{>}1$, we used open symbols. The data points with error bars are from simulations where we averaged over five runs with varying random seed. The Solar System bodies and `Oumuamua are shown with open diamond symbols and different colours: 1P/Halley (blue), 8P/Tuttle (orange), 9P/Tempel~1 (green), 19P/Borrelly (red), 67P/Churyumov-Gerasimenko (purple), 81P/Wild~2 (brown), 103P/Hartley~2 (pink), Arrokoth (grey), 1I/2017 U1 (`Oumuamua) (olive). \textbf{Left}: Axes ratios of the whole bodies. \textbf{Right}: Axes ratios of the individual lobes of the bilobed and contact-binary comets, Arrokoth, and the simulated planetesimals. We plot the axes ratios of the individual lobes with an upper triangle for the big lobe and a lower triangle for the small lobe. For Hartley~2, we included the two lobes that we found from analysing the shape model (see Appendix~\ref{sec:fitting103P}).}
\label{fig:aspect_ratio_map}
\end{figure*}

Figure~\ref{fig:aspect_ratio_map} visualises the axes ratios of Solar System bodies, `Oumuamua, and the planetesimals formed in our simulations. It is clear from this comparison that most of the Solar System bodies have axes ratios different to what we find in the simulations. The planetesimals formed in our simulations are very close to oblate spheroids with $b{/}a{\approx}1$ and $c{/}a{\lesssim}1$. The higher the angular momentum of the pebble cloud, the more flattened are the planetesimals. The disc-like shape of `Oumuamua would be consistent with the very flat planetesimals that form in pebble clouds with $1{\lesssim}L{/}L_\mathrm{J}{\lesssim}1.5$. On the other hand, most comets are prolate in shape or triaxial ellipsoids with $c{\approx}b{\lesssim}a$ (see also Table~\ref{tab:cometandKBOsizes}). Arrokoth as a whole has different axes ratios. Therefore, the observed shapes do not seem to be in agreement with the shapes of planetesimals formed by gravitational collapse. However, we neglected the fact that four out of the six comets that were visited by spacecraft are bilobed (19P/Borrelly, 67P/C-G, 103P/Hartley~2) or potentially bilobed (1P/Halley). Among the bilobed comets, 67P is a contact binary with two distinct lobes for which shape information is available \citep{Sierks2015,Jorda2016}. Furthermore, comet 8P/Tuttle is a contact binary as confirmed by radar observations \citep{Harmon2010}. And lastly, the cold classical Kuiper belt object Arrokoth is a contact binary as well with well characterised shape \citep{Keane2022}. Comets 19P/Borrelly and 103P/Hartley~2 are bilobed objects that might be the result of the formation from two individual objects \citep{Britt2004,Oberst2004,Thomas2013}. Using the available shape model for 103P/Hartley~2 \citep{Farnham2013}, we fitted ellipsoids to the two lobes of the comet (see Appendix~\ref{sec:fitting103P}). Comet 19P/Borrelly lacks a detailed shape model and we could not extract the two lobes. For 1P/Halley it is not entirely clear whether the nucleus is bilobed or not.

As outlined in Sect.~\ref{sec:introduction}, planetesimals are thought to form by gravitational collapse of pebble clouds. The observed properties of comets and the characteristics of Kuiper belt objects, for example the size distribution and the occurrence of binaries, are consistent with this formation hypothesis \citep{Blum2014,Blum2017,Blum2022,Nesvorny2010,Nesvorny2019,Nesvorny2021,Polak2023}. The dynamics of the collapse has been studied in detail with $N$-body simulations and shows that the pebble cloud typically fragments into a number of sub-clumps that eventually form planetesimals and bound binary systems \citep{Nesvorny2010,Robinson2020,Nesvorny2021,Polak2023}. The subsequent collapse of the binary or low-velocity collisions of two planetesimals would lead to the formation of contact binaries, such as Arrokoth or 67P. In this framework, the lobes of the contact binaries would be planetesimals that formed through gravitational collapse of the sub-clumps. Therefore, it is more reasonable to compare the shapes of the planetesimals in our simulations with the shapes of the lobes of the contact binaries. The right panel of Fig.~\ref{fig:aspect_ratio_map} shows how the simulated planetesimals compare to the lobes of the contact binaries. Besides the large lobes of 8P and 103P, the axes ratios of the other bodies overlap with the simulated planetesimals. This strongly supports the notion that the planetesimals forming a contact binary form from the collapse of a pebble cloud where angular momentum and gravity create oblate spheroids.

\subsection{The uncertain origin of 1I/2017 U1 (`Oumuamua)}
While the formation and nature of `Oumuamua is still under debate, a natural origin is advocated \citep{ISSI2019}. Different formation hypotheses are discussed in the literature that build on the observations of `Oumuamua, such as the unusually elongated shape \citep{Jewitt2017,Meech2017}, the lack of visible cometary activity \citep{Jewitt2017,Meech2017,Trilling2018}, and the non-gravitational acceleration \citep{Micheli2018}. It is unlikely that the non-gravitational acceleration is caused by strong volatile outgassing, as it is the case for Solar System comets, because the outgassing torques would spin up `Oumuamua to rotational fission \citep{Rafikov2018b}.

\citet{Flekkoy2019} discuss the possibility of  `Oumuamua being a fractal dust aggregate, which would be consistent with their estimates of the mechanical stability and the change in rotation period due to radiation pressure. \citet{Seligman2020} hypothesise that `Oumuamua could be a body rich in molecular hydrogen ice that formed in a cold dense core of a giant molecular cloud and that the sublimation of H$_2$ could explain the non-gravitational acceleration. \citet{Desch2021}, \citet{Jackson2021}, and \citet{Desch2022} argue that `Oumuamua could be a nitrogen-ice fragment excavated from an exo-Pluto with its shape heavily affected by N$_2$-ice sublimation while approaching the Sun. \citet{Raymond2018} use dynamical simulations to show that `Oumuamua could be an extinct fragment of an ejected cometary planetesimal. \citet{Bergner2023} interpret `Oumuamua as an icy comet-like planetesimal. They show that the non-gravitational acceleration of `Oumuamua could be driven by the outgassing of molecular hydrogen that was created by cosmic rays, trapped in the water ice matrix, and released upon annealing of the amorphous water ice matrix while approaching the Sun. \citet{Farnocchia2023} and \citet{Seligman2023} report the detection of sub-kilometre-sized near-Earth asteroids that exhibit significant non-gravitational acceleration that is greater than what can be attributed to the Yarkovsky effect. Similar to `Omuamua, these objects do not show any sign of visible coma, but weak volatile outgassing could explain their orbital motion.

In summary, despite the different hypotheses, the origin of `Oumuamua remains elusive. Being agnostic about the actual origin of `Oumuamua, our results demonstrate that the disc-like flattened shape of `Oumuamua could arise naturally from the gravitational collapse of a pebble cloud with high angular momentum.

\subsection{Limitations of the method}

Our method is capable of resolving the shapes of planetesimals formed by gravitational collapse. However, the fixed grid that is necessary for resolving pebble collisions sets some limitations for our model. In order to resolve the substructure of kilometre-sized planetesimal, we cannot follow the entire collapse from a Hill-sized pebble cloud to a planetesimal as it would require too many grid cells. Therefore, we are limited to follow the final assembly of a planetesimal from sub-clumps that form during the collapse \citep{Nesvorny2021,Polak2023}. Because of the fixed grid, cells that are filled to the maximum volume-filling factor will not move any more. Therefore, we cannot properly resolve the rotational state of the final planetesimal. A similar problem arises for pebble clouds with high angular momentum that collapse into disc-like flat structures ($L{/}L_\mathrm{J}{\gtrsim}1$). Once the disc structure has formed, the dense cells are frozen out. In reality, gravity and angular momentum would cause the substructures to collapse and form binaries or contact binaries. Therefore, our method is reliable for low angular-momentum clouds that collapse into a single body.


\section{Conclusion}
\label{sec:conclusions}

We developed a Monte-Carlo method to simulate the collapse of a rotating pebble cloud and the final assembly of a planetesimal. We investigated the shapes by analysing the axes ratios of ellipsoids fitted to the final body and compared our results to the shapes of Solar System comets with known shapes, the interstellar asteroid `Oumuamua, and the cold classical Kuiper belt object Arrokoth. We find that planetesimals that form by gravitational collapse of pebble clouds contain about $60{-}80\,\%$ of the mass of the cloud and ${\sim}50\,\%$ of the initial angular momentum. Furthermore, planetesimals that form from pebble clouds with initial angular momentum in the range $0.5{\lesssim}L{/}L_\mathrm{J}{\lesssim}1$, where $L_\mathrm{J}$ is the angular momentum of a critically rotating Jacobi ellipsoid with the same mass and effective radius as the planetesimal, match the observed shapes of the lobes of Arrokoth. The shapes of the lobes of bilobed comets agree with the shapes of the planetesimals that form from $L{/}L_\mathrm{J}{\lesssim}1$. Lastly, the disc-like shape of the interstellar asteroid `Oumuamua, if this body is indeed primordial and not a collisional fragment, is consistent with a formation in pebble clouds with $1{\lesssim}L{/}L_\mathrm{J}{\lesssim}1.5$. Because comets and, especially, Arrokoth most likely represent the primordial planetesimals of the Solar System (even though evolutionary processes might have lead to some reshaping), our results suggest that the flattened shapes of planetesimals naturally follows from the gravitational collapse of a rotating pebble cloud. This further strengthens the hypothesis of planetesimal formation by gravitational collapse.


\begin{acknowledgements}

We thank the anonymous referee for helpful comments that helped improving this manuscript. A.J. acknowledges funding from the European Research Foundation (ERC Consolidator Grant 724687-PLANETESYS), the Knut and Alice Wallenberg Foundation (Wallenberg Scholar Grant 2019.0442), the Swedish Research Council (Project Grant 2018-04867), the Danish National Research Foundation (DNRF Chair Grant DNRF159) and the Göran Gustafsson Foundation.

\end{acknowledgements}

\bibliographystyle{aa} 
\bibliography{ref} 

\begin{appendix}

\section{Monte Carlo method for solving the Enskog equation}
\label{sec:montecarlomethod}

The Monte Carlo method for solving the Enskog equation follows the procedure outlined in \citet{Montanero1997} and \citet{Poeschel2005}. We use a total of $N$ particles to sample the distribution function of the pebbles. Each particle represents a swarm of $N_\mathrm{swm}{=}N_\mathrm{phys}/N$ pebbles, where $N_\mathrm{phys}{=}M_\mathrm{p}/m_\mathrm{peb}$ is the total number of pebbles of mass $m_\mathrm{peb}$ that is needed for a planetesimal of mass $M_\mathrm{p}$. Each particle has position $\mathbf{x}$, velocity $\mathbf{v}$, mass $m$, and radius $R$. The Monte-Carlo method calculates the change of particle velocities owing to collisions. 

The number of collisions of particle $i$ in cell $I$ with particles $j$ in cell $J$ in a time step $\Delta t$ is
\begin{equation}
\omega_{ij}{=}\sigma^24\pi(\mathbf{v}_{ij}\cdot\mathbf{e}_i)\chi(\mathbf{x}_i,\mathbf{x}_i{+}\sigma\mathbf{e}_i)n_J\Delta t.
\end{equation}
Here, $\sigma$ is the diameter of a pebble in case of a mono-disperse size distribution of pebbles; for a non-uniform size distribution, $\sigma$ is chosen at random in the interval between the minimum and the maximum pebble radius such that $R_i{+}R_\mathrm{min}{\leq}\sigma{\leq}R_i{+}R_\mathrm{max}$. The relative velocity between particles $i$ and $j$ is $\mathbf{v}_{ij}{=}\mathbf{v}_j{-}\mathbf{v}_i$ and  $\mathbf{e}_i$ is the direction of the collision, that is, the unit vector pointing from particle $i$ to $j$. Finally, $\chi(\mathbf{x}_i,\mathbf{x}_i{+}\sigma\mathbf{e}_i)$ is the local equilibrium pair correlation function (or Enskog factor) \citep{VanBeijeren1973}, which depends on the density in the cell $J$, and $n_J$ is the number density in cell $J$. The local equilibrium pair correlation function can be determined from the equation of state \citep{Frezzotti1998}. Here, we use the hard-sphere equation of state from \citet{Ma1986}, which gives
\begin{equation}
\chi=\frac{1+2.5\phi+4.5904\phi^2+4.515439\phi^3}{(1-(\phi/\phi_\mathrm{max})^3)^{0.67802}}
\end{equation}
for the correlation function, where $\phi$ is the packing fraction and $\phi_\mathrm{max}$ is the maximum packing fraction. The maximum packing fraction, we set to $0.6$, which lies in the range between random loose packing (RLP) with $\phi_\mathrm{max}{=}0.56$ and random close packing (RCP) with $\phi_\mathrm{max}{=}0.64$ \citep{Onoda1990}.

The idea of the Monte Carlo method is to sample a certain number of collision for each cell. How many collisions are expected is calculated as
\begin{equation}
N_\mathrm{coll}{=}\tfrac{1}{2}N_I\omega_\mathrm{max},
\end{equation}
where $\omega_\mathrm{max}{=}\max_{ij}(\{\omega_{ij}\})$, for any particle $i$ of cell $I$ and any particle $j$ of cell $J$ separated a distance $\sigma$ from $I$, is an upper bound on the number of collisions \citep{Montanero1997}. For each collision, we then carry out the following steps:
\begin{enumerate}
\item A particle $i$ from all particles in the current cell $I$ is chosen at random with equiprobability.
\item A direction $\mathbf{e}_i$ is chosen at random with equiprobability.
\item A particle $j$ from the cell $J$ where $\mathbf{x}_i{+}\sigma\mathbf{e}_i$ points to is chosen at random.
\item The collision between particles $i$ and $j$ is accepted with probability $\Theta(\mathbf{e}_i\cdot\mathbf{v}_{ij})\omega_{ij}/\omega_\mathrm{max}$. Here, $\Theta(\mathbf{e}_i\cdot\mathbf{v}_{ij})$ makes sure that the particle pair is approaching.
\item If the collision is accepted, new velocities are assigned to particles $i$ and $j$ according to $\mathbf{v}_i^\prime{=}\mathbf{v}_i{+}(1{+}\epsilon)\frac{m_j}{m_i{+}m_j}\mathbf{v}_{ij}\cdot\mathbf{e}_i$ and $\mathbf{v}_j^\prime{=}\mathbf{v}_j{-}(1{+}\epsilon)\frac{m_i}{m_i{+}m_j}\mathbf{v}_{ij}\cdot\mathbf{e}_i$, where $\epsilon$ is the coefficient of restitution.
\end{enumerate}
Once all collisions are sampled, we proceed with the next cell.

Even though the Enskog equation takes excluded volume and correlations between particles into account, it does not prevent cells from becoming unrealistically dense, that is, having $\phi{>}\phi_\mathrm{max}$. The reason for this is that when particles are propagated, their new positions are unconditionally accepted \citep{Hong2021}. To prevent this from happening, new particle positions (if the particle enters a different cell) are accepted with a probability of $\exp[-6(\chi(\phi_1)\phi_1-\chi(\phi_0)\phi_0]$ where $\phi_{0,1}$ and $\chi(\phi_{0,1})$ are the packing fractions and correlation functions before and after the particle would enter the cell \citep{Hong2021}.

\section{Self-gravity of the cloud}
\label{sec:selfgravityofthecloud}

The Poisson equation,
\begin{equation}
\Delta \Phi(\mathbf{x}) = 4\pi\rho(\mathbf{x}),
\end{equation}
determines the dynamical evolution of the pebble cloud under its own gravity. Here, $\Phi$ is the gravitational potential and $\rho$ is the density of the pebble distribution. Poisson's equation can be easily solved in Fourier space. The potential and the density are written in terms of their Fourier transforms
\begin{align}
\Phi(\mathbf{x})&=\int\hat{\Phi}(\mathbf{k})e^{i2\pi\mathbf{k}\cdot\mathbf{x}}\,\mathrm{d}\mathbf{k}, \\
\rho(\mathbf{x})&=\int\hat{\rho}(\mathbf{k})e^{i2\pi\mathbf{k}\cdot\mathbf{x}}\,\mathrm{d}\mathbf{k},
\end{align}
where $\mathbf{k}$ is the wave number and a hat denotes quantities in Fourier space. Poisson's equation in Fourier space simplifies to the algebraic equation
\begin{equation}
\hat{\Phi}=-4\pi G\frac{\hat{\rho}}{4\pi^2|\mathbf{k}|^2}.
\end{equation}
We first calculate $\hat{\rho}$ using the Fast Fourier Transform (FFT), then solve the above equation, and in a last step calculate the inverse FFT of $\hat{\Phi}$ to obtain the potential $\Phi$ in real space. For the FFT, we make use of the FFTW package \citep{Frigo2005}, which is an highly optimised software package to carry out fast FFTs. We calculate the accelerations at the cell centres from the potential $\Phi$ using a standard central finite-difference scheme
\begin{align}
a_{x,ijk}&=-\frac{\Phi_{i+1jk}-\Phi_{i-1jk}}{2\Delta x}, \\
a_{y,ijk}&=-\frac{\Phi_{ij+1k}-\Phi_{ij-1k}}{2\Delta y}, \\
a_{z,ijk}&=-\frac{\Phi_{ijk+1}-\Phi_{ijk-1}}{2\Delta z},
\end{align}
where $\Delta x$, $\Delta y$, and $\Delta z$ are the grid spacings in each direction. Finally, we perform a linear interpolation to obtain the acceleration at the actual particle position.

The Fourier method implicitly assumes periodic boundary conditions. Therefore, it is important to chose a simulation box large enough for boundary effects owing to the periodic images of the pebble cloud to be unimportant. We found from testing different box sizes for the pure gravitational collapse of the pebble cloud that a box which is about two times larger than the pebble cloud diameter is sufficient.

\section{Fitting an ellipsoid}
\label{sec:ellipsoidalfit}

To determine the axes ratios of the planetesimals, we fitted an ellipsoid to the data points. This can be done as we describe in the following. The general equation for a conic section is given by the polynomial
\begin{equation}
Ax^2{+}By^2{+}Cz^2{+}Dxy{+}Exz{+}Fyz{+}Gx{+}Hy{+}Iz{+}J=0. \label{eq:conicsection}
\end{equation}
We have $N$ data points $(x_i,y_i,z_i)$ for the boundary of the planetesimal, which provides us with a set of $N$ equations of the above type of Eq.~\ref{eq:conicsection}. Normalising such that $J{=}{-}1$ and introducing the $N{\times}9$ matrix $M$ which has rows formed by the vectors $(
x_i^2 \; y_i^2 \; z_i^2 \; x_iy_i \; x_iz_i \; y_iz_i \; x_i \; y_i \; z_i)$, we can write the matrix equation
\begin{equation}
M\mathbf{p}=\mathbf{b}, \label{eq:ellipsematrix}
\end{equation}
where we have further introduced two vectors, one of length $9$ for the parameters $\mathbf{p}{=}(A \; B \; C \; D \; E \; F \; G \; H \; I)^\top$ and one of length $N$ for the right-hand side $\mathbf{b}{=}(1\dots1)$. The parameters of the ellipsoid are found by solving the above matrix equation~\ref{eq:ellipsematrix}. Because $M$ is not a square matrix, we multiply with $M^\top$ from the left side. The product $M^\top M$ is a square matrix which can be inverted to solve for $\mathbf{p}$, which we obtain through
\begin{equation}
\mathbf{p}=(M^\top M)^{-1}(M^\top \mathbf{b}).
\end{equation}
Having the parameters, the next step is to extract the properties of the ellipsoid: shape, orientation, and centre. To do so, we first notice that the polynomial in Eq.~\ref{eq:conicsection} can be written in matrix form as a quadratic form,
\begin{equation}
\mathbf{y}^\top Q\mathbf{y}=0,
\end{equation}
where we introduced the vector $\mathbf{y}{=}(x\,y\,z\,1)^\top$ and the matrix Q
\begin{equation}
Q=
\begin{pmatrix}
A & D/2 & E/2 & G/2 \\
D/2 & B & F/2 & H/2 \\
E/2 & F/2 & C & I/2 \\
G/2 & H/2 & I/2 & J 
\end{pmatrix}.
\end{equation}
To find the centre $\mathbf{x}_0{=}(x_0\,y_0\,z_0)$ of the ellipsoid, we introduce a translation matrix of the form
\begin{equation}
T=
\begin{pmatrix}
1 & 0 & 0 & x_0 \\
0 & 1 & 0 & y_0 \\
0 & 0 & 1 & z_0 \\
0 & 0 & 0 & 1
\end{pmatrix}
\end{equation}
such that $T\mathbf{y}$ moves the ellipsoid to the centre $\mathbf{x}_0$. Doing so, we can write the shifted ellipsoid in matrix form as $\mathbf{y}^\top(T^\top QT)\mathbf{y}$. The transformed matrix is symmetric and looks as follows
\begin{equation}
T^\top QT=
\begin{pmatrix}
A & D/2 & E/2 & T_{14} \\
D/2 & B & F/2 & T_{24} \\
E/2 & F/2 & C & T_{34} \\
T_{14} & T_{24} & T_{34} & T_{44} 
\end{pmatrix},
\end{equation}
where we notice that the upper left $3{\times}3$ matrix is the same as in $Q$. This is the part that describes the shape and orientation of the ellipsoid. The entries $T_{14}$ to $T_{34}$ arise from the fact that the ellipsoid is not centred at $(0,0,0)$ but shifted to $\mathbf{x}_0$, that is, the terms linear in $(x,y,z)$ in the polynomial form of the ellipsoid Eq.~\ref{eq:conicsection} do not vanish. And finally, the entry $T_{44}$ is the normalisation for recovering the enforced $J{=}{-}1$. The individual entries are
\begin{align}
T_{14}&=Ax_0+\frac{D}{2}y_0+\frac{E}{2}z_0+\frac{G}{2}, \\
T_{24}&=\frac{D}{2}x_0+By_0+\frac{F}{2}z_0+\frac{H}{2}, \\
T_{34}&=\frac{E}{2}x_0+\frac{F}{2}y_0+Cz_0+\frac{I}{2}, \\
T_{44}&=Ax_0^2+By_0^2+Cz_0^2+Dx_0y_0 \nonumber \\
&+Ex_0z_0+Fy_0z_0+Gx_0+Hy_0+Iz_0+J.
\end{align}
To find the centre of the ellipsoid, we notice that the entries $T_{14}{=}T_{24}{=}T_{34}{=}0$ need to be equal to zero. This gives the following linear matrix equation
\begin{equation}
\begin{pmatrix}
A & D/2 & E/2 \\
D/2 & B & F/2 \\
E/2 & F/2 & C
\end{pmatrix}
\begin{pmatrix}
x_0 \\
y_0 \\
z_0 \\
\end{pmatrix}
=
\begin{pmatrix}
-G/2 \\
-H/2 \\
-I/2
\end{pmatrix},
\end{equation}
which can be easily solved to find the centre. Next we want to find the axes and orientation of the ellipsoid. We can do this by finding the eigenvalues and eigenvectors of the matrix
\begin{equation}
\mathcal{E}=
\begin{pmatrix}
A & D/2 & E/2 \\
D/2 & B & F/2 \\
E/2 & F/2 & C
\end{pmatrix},
\end{equation}
which we need to normalise by dividing all entries by $-T_{44}$ in order to recover $J{=}{-}1$. The eigenvalues are the solution of the characteristic polynomial
\begin{equation}
\det(\mathcal{E}-\lambda\mathbb{I}_3)=0,
\end{equation}
where $\mathbb{I}_3$ is the $3{\times}3$ identity matrix and $\lambda$ is the eigenvalue. Because in diagonal form the matrix describing the ellipsoid is
\begin{equation}
\mathcal{E}_\mathrm{diag}=
\begin{pmatrix}
\lambda_{a} & 0 & 0 \\
0 & \lambda_{b} & 0 \\
0 & 0 & \lambda_{c}
\end{pmatrix},
\end{equation}
which results in $\lambda_{a}x^2{+}\lambda_{b}y^2{+}\lambda_{c}z^2{=}1$, the axes are related to the eigenvalues as $a{=}\lambda_{a}^{-1/2}$, $b{=}\lambda_{b}^{-1/2}$, and $c{=}\lambda_{c}^{-1/2}$. For each eigenvalue, a eigenvector $\mathbf{q}$ can be calculated by the standard procedure of solving the eigenvalue equation $\mathcal{E}\mathbf{q}{=}\lambda\mathbf{q}$, which gives the direction of the corresponding axis.

\section{Fitting two lobes to 103P/Hartley 2}
\label{sec:fitting103P}

\begin{table*}
\caption{Ellipsoidal fit 103P/Hartley 2.}
\label{tab:ellipsoidalfit103P}
\centering
\begin{tabular}{lcccccc}
\hline\hline
& $a$ & $b$ & $c$ & $x_0$ & $y_0$ & $z_0$ \\
\hline
full & $2.661$ & $0.830$ & $0.730$ & $-0.0072$ & $-0.0053$ & $-0.0577$ \\
big lobe & $1.604$ & $0.890$ & $0.777$ & $-0.0055$ & $-0.0053$ & $-0.3039$ \\
small lobe & $0.892$ & $0.737$ & $0.696$ & $-2.3{\times}10^{-3}$ & $-4.53{\times}10^{-4}$ & $0.8150$ \\
\hline
\end{tabular}
\end{table*}

Based on the shape model of 103P/Hartley~2 \citep{Farnham2013}, we cut the nucleus at the neck region in two parts. To do so, we identified the thinnest cross section of the neck and fitted a plane that separated the nucleus in a big and a small lobe, in a similar fashion as \citet{Keane2022} described it for Arrokoth. After that, we fitted ellipsoids to each lobe separately. The separation plane is given as $-0.0088x{+}0.0866y{+}z{=}{-}0.4161$, where $(x,y,z)$ are planetocentric Cartesian coordinates. The fitting results are summarised in Table~\ref{tab:ellipsoidalfit103P} and we use our results for the big and small lobes in Fig.~\ref{fig:aspect_ratio_map}.

\section{3D visualisation of a planetesimal}

\begin{figure}
\begin{center}
\resizebox{\hsize}{!}{\includegraphics{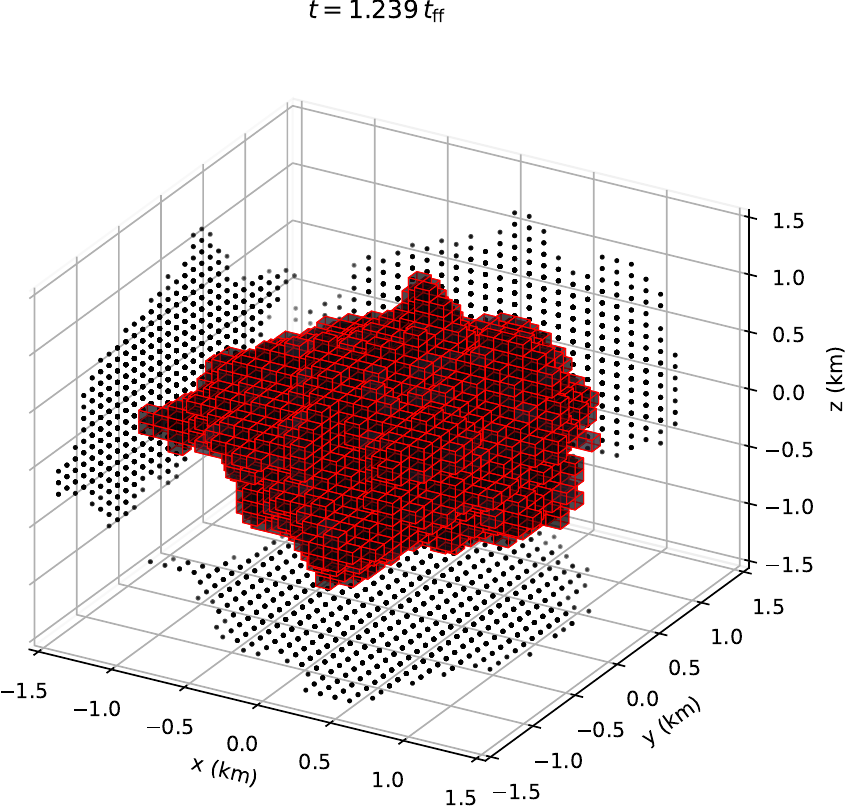}}
\end{center}
\caption{Shape model of a planetesimal with $L/L_\mathrm{J}{=}0.5$. Each volume element with a volume-filling factor ${\ge}0.5$ is plotted to visualise the shape of the planetesimal. The centres of the volume elements are projected into the $xy$-, $xz$-, and $yz$-planes to visualise the shape of the planetesimal that can not be seen in the three-dimensional representation.}
\label{fig:shape}
\end{figure}

As an example, Fig.~\ref{fig:shape} shows a 3D visualisation of a planetesimal formed in our simulations. The initial angular momentum of the cloud was $L/L_\mathrm{J}{=}0.5$. One can see the flattened structure of the planetesimal arising from angular momentum conservation.

\end{appendix}

\end{document}